\documentclass{article}

\newcommand{\be}{\begin{equation}}
\newcommand{\ee}{\end{equation}}
\begin{document}

\title{Topological Effects in Matrix Models
representing Lattice Gauge Theories at Large N
\thanks{Talk delivered at TH-2002, International 
Conference on Theoretical
Physics, Paris, UNESCO, July 22-27, 2002.}
}

\author{H. Neuberger\\~\\
School of Natural Sciences, \\ Institute for Advanced Study, 
Princeton, NJ 08540\\~\\
Department of Physics and Astronomy\thanks{Permanent address.},\\ Rutgers University,
Piscataway, NJ 08855\\~\\
{\sl neuberg@physics.rutgers.edu}\\}

\maketitle

\begin{abstract}
Quenched reduction is revisited from the modern viewpoint 
of field-orbifolding. Fermions are included and it is shown
how the old problem of preserving anomalies and field topology
after reduction is solved with the help of the overlap construction.
\end{abstract}

\section{Introduction}

The techniques producing numerical values for physical
quantities affected by the strong interactions have
continuously  evolved over the last twenty years.
They have become quite sophisticated and, coupled with
major increases in affordable computing power, generate steady, 
incremental progress~\cite{Lurcher}. However, if the sole activity in 
lattice field theory were
that of adding digits of higher accuracy to some 
basic quantities one might get the impression that 
this subject, while useful, is rather dull.  
Luckily, this is not the case: there are lattice 
activities which do take risks and, at times, produce a breakthrough
(e.g. lattice chiral fermions~\cite{overlap}), 
or, show that lattice field theory is
good for other things than QCD (e.g. Higgs triviality bounds
~\cite{Higgs}). There are
people who venture into supersymmetric territory
~\cite{susy}, tackle ``insoluble
problems'' like QCD with a chemical potential~\cite{fodor}, 
use lattice techniques
to check scenarios for explaining the baryon number 
asymmetry~\cite{shapo}, 
and are inventive in many other ways. In turn, lattice 
field theory constructions enter mainstream particle physics
~\cite{deconstr} and even quantum gravity~\cite{grav-deconstr}, 
providing novel ``UV-completions''.

In this talk I shall present a project whose long term
objective is to solve 4D planar QCD. The project is in its initial
stages and, unlike more typical lattice 
field theory projects, is far from
being guaranteed to succeed. This project 
is a collaborative effort with J. Kiskis and R. Narayanan 
~\cite{ourproject}. 
At present, the focus is narrowed down to
the calculation of the $Q^2$-dependent meson-meson 
correlation functions in 4D. In 2D, where the planar limit is
solvable (as shown by 't Hooft~\cite{thooft2d}), this correlation
function is of fundamental importance. 
Although 't Hooft's solution was mainly
analytical, to actually get numbers also a numerical procedure
is required, even in two dimensions. 
There is little chance for a similar analytical breakthrough
in 4D. But, we hope that a numerical approach - this time containing
a stochastic element - will be possible using old ideas of large $N_c$
reduction and new lattice fermions. 

't Hooft's planar limit~\cite{thooft} 
in 4D requires three separate limits, one more
than in ordinary QCD. One needs to take an ultraviolet cutoff, $\Lambda$,
(needed to define the theory beyond its perturbative series)
to infinity, an infrared cutoff (typically the 
volume of an Euclidean torus), V, to infinity and the number of colors,
$N_c$, at $g^2N_c=\lambda$ fixed, also to infinity ($\lambda$ is to be traded
for a physical scale). 
The natural order of limits is as follows: first $V\to\infty$, next
$\Lambda\to\infty$ and last $N_c\to\infty$. One knows how to exchange
the first two limits without altering the result 
even when the quarks are massless. 
Less is known about taking $N_c\to\infty$ earlier than last, but
we shall do that. Otherwise, we are relegated to solving not only QCD
first, but also QCD-like theories for different numbers of colors,
and then the relevance of getting the planar limit numerically 
becomes mostly academic. The result~\cite{teper} is nevertheless important for
our project because such investigations tell us, quantitatively, how far
$N_c=\infty$ is from $N_c=3$. 
For us,  ``solving'' planar QCD means that a shortcut 
to the $N_c=\infty$ limit was used, and the method 
finds a numerical solution at $N_c=\infty$ 
with considerable more ease than would be possible in real QCD. 
This ancient hope sounds more realistic today than it did five years
ago also thanks to progress made by string theorists~\cite{bigrev}. 
To be sure, in general, the interchange of the limits $N_c\to\infty$ 
with $\Lambda\to\infty$ is non-trivial~\cite{neub}. For example, the theory may
contain a set of degrees of freedom whose number does not increase
with $N_c$, but whose impact on the ultraviolet behavior at any
finite $N_c$ is large.
Nor is the interchange of
the $V\to\infty$ and $N_c\to\infty$ limits a trivial matter in a theory
with massless elementary fields~\cite{thooft2}.

\section{Large $N_c$ reduction} 

We start on a lattice with a ``bare'' coupling $\beta=\frac{4}{g^2}$
and at a finite volume $V$. There is very good evidence that the limit
$N_c\to\infty$, with $b\equiv\frac{\beta}{N_c}$ fixed, exists. It exists
for the entire range of positive values of $b$. At small values of $b$
(strong coupling) large $N_c$ reduction 
~\cite{ek}
tells us that the $V\to\infty$
limit is trivial: there is no $V$-dependence left at leading order in
$N_c$; one can set $V=1$ and lattice momentum space is generated from
the eigenvalues of the gauge field operators. Complications arise when
one tries to extend this to weak coupling (large $b$)
~\cite{bhn}, but there are some
options available~\cite{bhn}. This is not dissimilar to the situation in models
that admit a weakly coupled dual string description at large $\lambda$. 

In their work, Eguchi and Kawai~\cite{ek} proved that the Schwinger-Dyson
equations for single-trace operators in the reduced and full models
were the same at infinite $N_c$. Later~\cite{bhn} it was realized that this
only established reduction for small enough $b$, and that to
extend it to large $b$ an additional ingredient, for example, 
``quenching''~\cite{bhn}, was needed.

In modern language, large $N_c$ reduction works by ``field orbifolding''
~\cite{f-o}.
This terminology comes from continuum field theories (actually
these theories are the low energy sector of some string theories and the
term originates from a method of constructing special backgrounds for
the latter~\cite{kachru}), but the dependence
on continuous space-time plays little role in the basic combinatorial
identity. In our lattice application there is no explicit continuum.
Field orbifolding can produce pairs of theories that admit 't Hooft
expansions and, at least at the level of bare diagrams, are simply
related by possible rescalings of the coupling constant. The fields
in one member of the pair are obtained by subjecting the fields
in the other member to a set of constraints. These constraints
reduce the symmetry group from $U(N_c)\times {\bf G}$ to 
$\otimes_i U(N^i_c)\times {\bf G^\prime}$. 
In the context of large $N_c$ reduction
one starts with a theory that has $V=1$, a single site lattice, 
and one ends up with a
theory with a larger lattice 
(here chosen as a four dimensional symmetrical torus), $V=L^4$. 
The original theory has a $U(N_c)\times U(1)^4$ 
and the new theory has a symmetry
$\left [ \otimes_{i=1,L^4} U(N_c/L^4)\right ]\times U(1)^4\times Z(L)^4$. 
(The $N^i_c = N_c/L^4$ and are assumed to be integers.)
The two theories have the same $b$ and, in the second,  
the factor $\otimes_{i=1,L^4} U(N_c/L^4)$ is a lattice gauge symmetry group. 
The original theory is the reduced model and the main assumption
of our project is the hope that
the factor $L^4$ isn't really needed to get to the planar limit.

Let me first illustrate the main trick in a simpler
model, where the field integration measure, excepting the
action term, is flat and unrestricted. This makes it easy
to implement the constraint on the original fields 
by a ``decoration'' of the original planar diagrams.
The ``decoration'' is a discrete gauge-field theory 
whose gauge link variables need to be
summed over, living on the planar diagram of the
original theory. 

The model is known as the ``Weingarten model''~\cite{weingart}, and it
defines a generating functional for random surfaces on
a hypercubic lattice - a rudimentary string theory. 
The partition function of its reduced
version~\cite{ekw} is given by
\begin{equation}
Z^N_{\rm DW}(b) =\prod_{\mu=1}^d\prod_{1\le i,j\le N}
\left [ \int_{-\infty}^{\infty} 
dA^\mu_{ij} dA^{\mu*}_{ij} \right ]
e^{\frac{b}{N}\sum_{\mu\ne\nu} tr A_\mu A_\nu A^\dagger_\mu A^\dagger_\nu
-\sum_\mu tr A_\mu A^\dagger_\mu}
\end{equation}
Although the integral does
not converge, the formula is meaningful as a compact 
definition of an infinite series in $b$. Similarly, the
large $N$ limit of $F(b)\equiv\frac{1}{N^2}\log (Z^N_{\rm DW}(b))$ is
also meaningful as a series in $b$. 
It is given there by the sum of all planar $|\Phi|^4$-type
Feynman diagrams with a factor of $b$ attached to each vertex. 

The reduced model has a symmetry $SU(N)\times U(1)^d$, where
the $SU(N)$ acts by conjugation simultaneously on all of 
the $A_\mu$ matrices and the
$U(1)$ factors independently affect the phases of each $A_\mu$
matrix. 
There also is a symmetry group made out of $2^d d!$ elements
consisting of permutations of the $\mu$ indices and conjugations.
This ``crystallographic group'' plays no role in our particular
application here, and hence will be ignored (it also was ignored
a few paragraphs earlier). 
The constrained, or, equivalently, space-time-extended model,  
is defined on a
hypercubic toroidal lattice containing $L^d$ sites. Its
partition functions is:
\begin{eqnarray}
Z^{N^\prime}_{\rm CDW}(b) =\prod_{\mu=1}^{d}\prod_{x_\mu =0}^{L-1}~
\prod_{1\le i,j\le N}\left [ \int_{-\infty}^{\infty} 
dA^\mu_{ij} (x) dA^{\mu*}_{ij} (x) \right ]\nonumber\\
e^{\frac{b^\prime}{N^\prime}\sum_{\mu\ne\nu} \sum_x 
tr A_\mu (x) A_\nu (x+\mu) A^\dagger_\mu (x+\nu) A^\dagger_\nu (x)
-\sum_\mu \sum_x tr A_\mu (x) A^\dagger_\mu (x)}
\end{eqnarray} 
The symmetry of the extended model 
is $\left [ \otimes_x U(N^\prime ) 
\right ] \times U(1)^d \times Z(L)^d$. 
The first factor is a local lattice gauge
symmetry, the second is a global symmetry and
the third a space-time symmetry. 
Again we ignore the crystallographic group. 
Actually, the phases
associated with the $U(1)$'s go only from zero to $\frac{2\pi}{L}$ -
larger phases can be reduced by lattice gauge transformations.
Again the integral does
not converge, but the formula is meaningful as a compact 
definition of an infinite series in $b^\prime$. Similarly, the
large $N^\prime$ limit of 
$F^\prime(b^\prime)\equiv\frac{1}{L^d (N^\prime )^2}
\log (Z^{N^\prime}_{\rm CDW}(b^\prime))$ is 
meaningful as a series in $b^\prime$. 
It is again given by the sum of all planar $|\Phi|^4$-type
Feynman diagrams with a factor of $b^\prime$ attached to each vertex,
but, with propagators that are less trivial than before. 

The extended model can be viewed as a constrained version of the reduced
model if we pick $N=N^\prime L^4$, $b=b^\prime$, write the $1\le i \le N$
indices as double indices $(i^\prime ,x)$ with $1\le i^\prime \le N^\prime$,
$0\le x_\mu \le L-1$ 
and restrict the $[A_\mu]_{(i^\prime, x),(j^\prime, y)}$ to 
be zero when $y\ne x+\hat\mu$. When $y=x+\hat\mu$ there are no
additional restrictions. If all $A_\mu$'s in the reduced model of
eq. (1) obey this constraint, the portion of the trace referring to
the $x,y$ components of the indices can be written out explicitly
and one obtains the extended model of equation (2). 
An important observation is that the constraint can be viewed as enforcing
invariance under a subgroup of the symmetry group of the reduced model:
\begin{eqnarray}
A_\mu = {\cal T_\nu} (A_\mu) = e^{-\frac{2\pi\imath}{L}\delta_{\mu,\nu}} 
g^{-1}_\nu A_\mu g_\nu, ~~~~~~~~~ \nu=1,..,d\\
\left [ g_\nu\right ]_{(i^\prime,x),(j^\prime,y)} = \delta_{i^\prime,j^\prime} 
\delta_{x,y} e^{\frac{2\pi\imath}{L}y_\nu}
\end{eqnarray}
The set of operations ${\cal T_\nu}$ generates a $Z(L)^d$ subgroup
of the symmetry group of the reduced model. The elements of the group
are labeled by $[s]$ where $[s]=(s_1,s_2,...,s_d)$ and $s_\mu=0,1,...,L-1$.
\begin{equation}
{\cal T}^{[s]} (A_\mu ) = e^{\frac{2\pi\imath}{L}s_\mu }
\left ( g^{[s]} \right )^\dagger A_\mu
g^{[s]},~~~~g^{[s]}=\prod_{\nu=1}^d g_\nu^{s_\nu}
\end{equation}
Note that the action of ${\cal T}^{[s]}$ on $A_\mu$ differs from its
action on $A_\nu$ for $\mu\ne\nu$ by an overall phase factor. 
This $Z(L)^d$ intersects
both original symmetry group factors, $SU(N)$ and $U(1)^d$. 
Invariance under the action of the four generators ${\cal T_\nu}$
is equivalent to invariance under the entire $Z(L)^d$ group. 

Since we integrate freely on the entries of the matrices $A_\mu$, we
can redefine the constrained model by keeping full matrices in the
quadratic term (where the constrained and unconstrained entries of
the $A_\mu$-matrices do not mix) 
and only matrices satisfying the constraint in the
quartic term. The constraint can be put in effect by 
first expanding in the exponential of the quartic factor, 
and then replacing
each occurrence of a constrained $A_\mu$, independently, 
by ${\cal T}^{[s]}(A_\mu)$ with an unconstrained $A_\mu$,  
and averaging 
each of the many ${\cal T}^{[s]}$ factors 
over the group $Z(L)^d$. 
Only the parts of $A_\mu$ that are invariant,
making up the matrix $A_{\mu ~{\rm constr}}$, survive
this group averaging, while the other entries are set
to zero. 
We now leave  
this group averaging to the end of the calculation, 
and, instead, do the Gaussian
integrals first, producing Feynman diagrams. The propagators are
now given by
\begin{equation}
< {\cal T}^{[t]}(A_\mu)_{ij} {\cal T}^{[s]} (A^\dagger_\nu)_{kl} > =
e^{\frac{2\pi\imath}{L} (t_\mu - s_\mu )} 
g^{[s-t]}_{il} g^{[t-s]}_{kj} \delta_{\mu\nu}
\end{equation}
${\cal T}^{[t]}$ and ${\cal T}^{[s]}$ appear only in this propagator. 
Moreover, as we see, there is only dependence on $[t-s]$.
Hence, the independent averages over $[t]$ and $[s]$
can be replaced by a single average over the combination
$[t-s]$. It is only this extra group averaging procedure 
that distinguishes the two models. 

The sum over the $i,j,...$ indices produces a factor of $N$
for each index loop in the double line notation for propagators
in the unconstrained model. In the extended model, 
because of the constraints, 
these index sums produce
a trace of the product of the various $g^{[s]}$ factors
encountered round the index loop. The $g^{[s]}$ matrices also form a group,
$Z(L)^d$, but this group is distinct from the $Z(L)^d$
made out of the ${\cal T}^{[s]}$ operations. Thus, for every index
loop we need to calculate the trace of a $g^{[t]}$ that is the parallel
transporter round the loop. 

{}From their definition we see that the $g^{[s]}$ obey:
\begin{equation}
tr g^{[s]} = N\delta_{[s],[0]}
\end{equation}
where the delta function forces the group element to be unity. 
This has the important consequence that the phase factors
that appear in the definition of the ${\cal T}^{[s]}$ operations
cancel out any time the contribution from the index loops is non-vanishing.
(Note that the phase factors are 
associated with propagators, not index lines.)
Thus, we can ignore the phase factors and 
take the propagators as given by
\begin{equation}
< {\cal T}^{[t]}(A_\mu)_{ij} {\cal T}^{[s]} (A^\dagger_\nu)_{kl} >
\rightarrow
g^{[u]}_{il} g^{[-u]}_{kj}
\end{equation}
Group averaging amounts to averaging over all $L^d$ values
of the $[u]$ labels.

More generally, this induces us to think about a situation in which
we assign to each propagator in the planar 
diagram a group element $g\in G$
where $G$ is some finite group. $g$ is represented by an $N\times N$
matrix $D_{ij}(g)$ which replaces one of the two index lines
making up the propagator. Before the ``decoration'', this index
line was a Kronecker delta. The other index line of the same
propagator is decorated by $D_{kl}(g^{-1})$. The group element
associated with each propagator is independently averaged over all of $G$
and the decoration introduces a weighting factor given by
the product over all index loops (faces) of traces of products of 
$D(g)$ matrices
round the loop. The summations over $G$ are now easy to perform -
a standard exercise in lattice strong coupling expansion,  
this time defined on a random two dimensional lattice~\cite{migdal}. 

For any $h\in G$, $tr D(h)$ is a class function and hence expandable as:
\begin{equation}
tr D(h) =\sum_r c_r d_r \chi_r (h)~~~~~
c_r=\sum_{h\in G} \chi_r(h^{-1}) \frac{1}{d_r} tr D(h)
\end{equation}
The averaging over the $g$'s associated to the propagators produce
a total factor given by
\begin{equation}
\sum_r c_r^f d_r^2 = \sum_r c_r^{2+v} d_r^2
\end{equation}
where $f$ is the number of faces and $d_r$ the dimension of the
$r$-irreducible representation of $G$. In the above equation 
we make use of the 
planarity of the diagram. If $e$ is the number of edges in the
diagram and $v$ the number of vertices, we have:
\begin{equation}
f-e+v=2~~~~~e=2v~~~~~~f=2+v
\end{equation}
In order to be able to absorb the decoration factor into a rescaling 
of $b^\prime$
we need $c_r=K$ for all $r$. {}From (9), $K=\frac{N}{d(G)}$, where
$d(G)$ is the number of elements in $G$. $d(r)$ is the dimensionality
of the irreducible representation $r$, and $d(G)=\sum_r d_r^2$. 
The contribution of the diagram is  
$\left ( \frac{b^\prime }{N^\prime }\right )^v\sum_r c_r^f d_r^2$. 
$K$ is an integer and we see that the representation of $G$ furnished by $D(h)$
contains $K$ copies of the regular representation. This is a consequence
of the requirement to have an overall $b^{\prime v}$ factor. 
It is easy to check that in our case, where $G=Z(L)^d$ (with elements
$g^{[s]}$), this condition on the representation holds. 

Turning now to our special case, we set $d(G)=L^d$, $N^\prime =
\frac{N}{d(G)}$ and $b^\prime=b$. 
The weight of the above diagram becomes 
$N^2 b^v \frac{1}{d(G)}$. 
{}From the definition of the free energy of the reduced and extended model
we now conclude that
\begin{equation}
F^\prime (b) = \frac{d(G)}{N^2} \log ( Z^{N^\prime}_{\rm CDW} (b)) = F(b)
\end{equation}
The equality of the free energies is the main result of reduction. 
Relations between other correlation functions can be obtained
by adding more terms to the action and taking derivatives with
respect to the new, infinitesimal, couplings. 
The above result was obtained a long time ago~\cite{ekw}  
by direct inspection of the series 
and by mapping the Schwinger-Dyson equations of single-trace
operators from one model to the other. 

In its extended version, the Weingarten model will turn into a 
lattice gauge theory if we constrain the $A_\mu (x)$ matrices to be
simple-unitary and change the integration measure to the $SU(N)$ 
group Haar measure. The relation between the reduced and extended
versions holds even after the unitarity constraint has been
imposed on both models. This was proven a long time ago, 
first by the Schwinger-Dyson
equation method~\cite{ek} and, subsequently, by inspection of lattice 
strong coupling diagrams~\cite{arroyo}  
and also using stochastic quantization~\cite{alfaro}.
A proof by field orbifolding would require 
a generalization of the analysis above, but since the
result is known to hold, there is little doubt that such a generalization
exists. Field orbifolding admits generalizations
involving nonabelian global symmetries, and it would be interesting to
see whether previously unknown results in planar lattice gauge theory
could be obtained from this approach. 

\section{Prototype model and its extended variants}

The unconstrained, reduced, gauge model consists of 
4 matrices $U_\mu \in SU(N_c )$, 
4 unitary matrices $V_\mu \in SU(N_f )$
and two rectangular
Grassmann matrices $\bar\Psi,\Psi$, where one index 
is in the fundamental representation of $SU(N_c)$ (color), and the other
index 
in the fundamental representation of $SU(N_f)$ (flavor). The fermion
matrices also have a Dirac index, which we often suppress. 
The model is defined by:
\begin{eqnarray}
S=\beta^u \sum_{\mu >\nu} 
C^u_{\mu\nu}C^{u\dagger}_{\mu\nu} +
\beta^v \sum_{\mu >\nu} 
C^v_{\mu\nu}C^{v\dagger}_{\mu\nu} +
Tr \left [\bar\Psi D(U,V) \left ( \Psi \right )\right ]\nonumber\\
C^u_{\mu\nu}\equiv[U_\mu ,U_\nu ],~~~
C^v_{\mu\nu}\equiv[V_\mu ,V_\nu ]
\end{eqnarray}
The linear operator $D(U,V)$ is a reduced version of a
lattice Dirac operator, to be introduced later on. 

The constrained associated model lives on an $L^4$ lattice~\cite{orland}.
The constraints mean that the matrices $U_\mu ,V_\mu ,\bar\Psi,\Psi$ 
are expressible in terms of $L^4$ smaller matrices
$u_\mu (x),v_\mu (x),\bar\psi(x),\psi(x)$. The dimensions 
are $n_c$ instead of $N_c$ and $n_f$ instead of $N_f$ where
$n_{c,f}=N_{c,f}/L^4$. ($n_{c,f},N_{c,f},L$ are all integers.)
\begin{eqnarray}
\left [ U_\mu \right ]_{(ix),(jy)} = [u_\mu (x)]_{ij} \delta_{y,x+\mu},
~~~\left [ V_\mu \right ]_{(ax),(by)} = [v_\mu (x)]_{ab} \delta_{y,x+\mu}
\nonumber\\
\left [ \Psi \right ]_{(ix),(ay)} = [\psi (x)]_{ia} \delta_{xy},~~~
\left [ \bar\Psi \right ]_{(ay),(ix)} = [\bar\psi (x)]_{ai} \delta_{yx}
\end{eqnarray}
For small lattice gauge coupling the constrained and unconstrained
models will be equivalent in the limit $N_{c,f}\to\infty$ with
$\frac{N_f}{N_c}$ held fixed and a simple $D(U,V)$. But, for larger coupling
the equivalence breaks down and we shall adopt here one of the
proposals for keeping the equivalence up even at large lattice couplings,
namely quenching~\cite{bhn}.

\section{Quenching}
Quenching amounts to freezing the eigenvalues of $U_\mu,V_\mu$ to
sets that are uniformly distributed round the circle. 
Other than that the unitary matrices are free. 
One also takes the rescaled coupling, 
$\frac{\beta_f}{N_c}$, to infinity so as to 
freeze out the flavor gauge
group~\cite{neublevine1}. 
The $U_\mu$ eigenvalues play the role of lattice momenta 
in boson propagator lines, while the $V_\mu$
eigenvalues enter only into the fermionic lines. The operator $D(U,V)$
breaks up into $N_f$ blocks of size $4N_c\times 4N_c$ each. This is the 
crucial point of our approach: The numerical price of 
handling fermions
has a chance to simplify tremendously if indeed $N_c$ does not
need to be anywhere nearly as big as the parameter 
$n_c L^4$ typical to numerical QCD
simulations and, nevertheless, the planar limit stays well approximated. 

\section{Overlap fermions}

We now define $D$ for one of the flavor blocks. So long we are focusing only
on fermions we can absorb 
the flavor phase into the color matrices $U_\mu$ (whose eigenvalues are
quenched as before). Define the unitary operators $X_\mu$:
\begin{equation}
X_\mu=\frac{1-\gamma_\mu}{2} U_\mu + \frac{1+\gamma_\mu}{2} U^\dagger_\mu
\end{equation}
The Wilson-Dirac matrix $D_W$ is a given by
\begin{equation}
D_W=m+\sum_\mu(1-X_\mu)
\end{equation}
We choose the parameter $m$ to be about $-1$ (it must be between $-2$ and $0$).
It is easy to see that $H_W\equiv \gamma_5 D_W$ is hermitian. 

The overlap construction says that the chiral determinant line bundle
is represented on the lattice by the inner product of two states, 
$\langle + | -\rangle$. Here $\langle +|$ is the eigen-bra of
a system of noninteracting second quantized fermions in a single body
potential given by $H_W$ with the mass sign switched, while
$|-\rangle$ is the eigen-ket of a system of noninteracting second 
quantized fermions in a single body
potential given by $H_W$ with the mass chosen as above. One can change
the lattice definition without affecting the continuum limit
so that the mass in $H_W$ is taken to be $+\infty$ for 
the $\langle +|$-state.
In that case the $H_W$ for the $\langle +|$-state can
be replaced by $\gamma_5$~\cite{overlap}. 

The overlap $\langle + | -\rangle$ can be written in terms of single
particle wave-functions as the determinant of a matrix $A$ defined 
below~\cite{overlap, narayanan}. Choose the chiral representation for 
the $\gamma_\mu$ matrices, so that
\begin{equation}
\gamma_5=\left ( \begin{array}{cc}1&0\\0&-1\end{array} \right )
\end{equation}
Let $\Phi$ be a simple unitary 
matrix of eigenvectors of $H_W$,  ordered so that
\begin{equation}
{\rm sign} (H_W) \Phi = \Phi 
\left ( \begin{array}{cc}1&0\\0&-1\end{array} \right )
,~~~~\Phi=\left ( \begin{array}{cc}A&B\\C&D\end{array}\right )
\end{equation}
Then 
\begin{equation}
\langle + | -\rangle=\det A
\end{equation}

For a vector-like theory we need $|\det(A)|^2$, which, unlike $\det A$, 
is a function, not a section of a bundle. Simple linear algebra
tells us that 
\begin{equation}
|\det (A) |^2=\det(A)\det(D)
\end{equation}
However, 
\begin{equation}
\gamma_5 {\rm sign}(H_W) \Phi = 
\left ( \begin{array}{cc}A&-B\\ -C & D\end{array} \right )
\end{equation}
Hence,
\begin{equation}
\det \left [ \frac{1+\gamma_5{\rm sign}{(H_W)}}{2} \Phi \right]=
\det \left [ \frac{1+\gamma_5{\rm sign}{(H_W)}}{2} \right]=
|\langle + | -\rangle|^2
\end{equation}
We discover that chiral symmetry can be preserved on the lattice if
we use the overlap Dirac operator~\cite{overlap}: 
\begin{equation}
D_o =\frac{1+\gamma_5{\rm sign}{H_W}}{2} 
\end{equation}

The operator $D$ is constructed out 
of $D_o$, with the addition of a small mass
term for the quarks. 

\section{Topology and Anomalies}

Anomalies are expressed by particular Berry-phase factors innate 
to the $|-\rangle$ state of the overlap~\cite{geom}. They obstruct 
smooth gauge invariant sections in cases where anomalies do not cancel.

Topology is expressed by $H_W$ having unequal numbers of positive and negative
eigenstates (above, where $A$ is taken as a square matrix, we 
tacitly assumed these two numbers to be equal
but this need not be so, in which case what we called
``$\det A$'' is identically zero). 
One can easily exhibit explicit sets of $U_\mu$
matrices for which there is non-zero topology. 

Problems arise if $H_W$ has an eigenvalue that is exactly zero. 
But this will
not happen as the continuum limit is approached because
one can prove the following inequality~\cite{bound}:
\begin{equation}
||[U_\mu,U_\nu]||\le \epsilon \Rightarrow 
\frac{\langle \psi | H^2_W |\psi\rangle}
{\langle \psi | \psi\rangle}
\ge 1 -|1+m|-\delta(\epsilon)
\end{equation}
where $\delta(\epsilon)$ goes to zero as $\epsilon$ goes to zero. 

At infinite $N_c$, for large enough lattice couplings and a
standard lattice action, 
the commutators indeed become bounded~\cite{ourproject}, as required. 

\section{Some history}

The first treatment of fermions in reduced models was carried out by
Gross and Kitazawa~\cite{grosskita} 
who invented the momentum ``force-feeding'' 
prescription that was derived above as a 
remnant of the frozen-out flavor gauge sector~\cite{neublevine1}.
These authors also invented a version of 
reduction that used hermitian matrices, $A_\mu$, 
rather than the unitary link matrices 
one has on a lattice~\cite{grosskita}. 
The Gross-Kitazawa model can be obtained from the lattice model 
by approximating $U_\mu\approx 1+\imath A_\mu$. 
However, as also noted by Gross and Kitazawa, 
this approximation does away with anomalies and topology. 
For many years we did not known how to cure this problem on the lattice,
where the link matrices are unitary. The problem was the same as the  
notorious problem of lattice chirality.
As we have seen, recent progress on the chiral 
fermion problem~\cite{overlap} 
makes it now possible to preserve all the necessary 
fermionic properties also in reduced lattice models removing
the main obstacle of principle to making nontrivial use
of reduction in numerical simulations.

\section{Acknowledgments}

I am grateful to Laurent Baulieu and to 
Daniel Iagolnitzer for the invitation to
TH-2002. I wish to acknowledge partial support 
at the Institute for Advanced Study
from a grant in aid from the Monell Foundation 
as well as partial support
by the DOE under grant DE-FG02-01ER41165 at Rutgers University.

\end{document}